\documentclass[aps,prl,reprint,preprintnumbers,slashbox]{revtex4-1}
\usepackage{latexsym,graphicx}
\usepackage{graphicx}
\usepackage{bm}
\usepackage{epsfig}
\usepackage{ulem}
\usepackage{color}
\usepackage{mathrsfs}
\usepackage{dcolumn}
\usepackage{setspace}
\usepackage{array}
\usepackage{amsmath}
\usepackage{amssymb}
\usepackage{dsfont}
\usepackage{gensymb}
\usepackage{dcolumn}
\usepackage{multirow}
\usepackage{bibentry,natbib}
\usepackage{booktabs}
\usepackage{latexsym}
\usepackage{diagbox}
\usepackage{epstopdf}
\usepackage{url}
\usepackage{hyperref}

\newcommand{\abs}[1]{\ensuremath{\left| #1 \right|}}

\newcolumntype{P}[1]{>{\centering\arraybackslash}m{#1}}

\begin{document}
\title{Black and White Holes at Material Junctions}

\author{Yaron Kedem${^1}$, Emil J. Bergholtz${^1}$, and Frank Wilczek${^1}{^2}{^3}{^4}$}
\affiliation{{}\\ $^1$Department of Physics, Stockholm University, Stockholm SE-106 91 Sweden\\
$^2$Center for Theoretical Physics, Massachusetts Institute of Technology, Cambridge, Massachusetts 02139 USA\\
$^3$T. D. Lee Institute, and Wilczek Quantum Center, Department of Physics and Astronomy, Shanghai Jiao Tong University, Shanghai 200240, China\\
$^4$Department of Physics and Origins Project, Arizona State University, Tempe AZ 25287 USA}
\begin{abstract}
Electrons in Type II Weyl semimetals display one-way propagation, which supports totally reflecting behavior at an endpoint, as one has for black hole horizons viewed from the inside.  Junctions of Type I and Type II lead to equations identical to what one has near black hole horizons, but the physical implications, we suggest, are quite different from expectations which are conventional in that context.  The time-reversed, ``white hole" configuration is also physically accessible.
\end{abstract}
\preprint{MIT-CTP/5165}
\maketitle

There has been much interest in the simulation of black hole properties by laboratory systems \cite{AnalogueGravity,deNovaObservation,DroriObservation,MomentumSign,TrappingLight,GravitationEffect,questions,Mechanism}.   Particularly intriguing, in this connection, is the possibility of realizing essentially quantum phenomena, notably including Hawking radiation \cite{hawking1975} or the closely related Unruh effect \cite{UnruhEffect}.    In the laboratory systems physical space is essentially flat, but the equations of motion for some degrees of freedom resemble those for particles near a black hole event horizon.   Systems under consideration include ``sonic black holes'' \cite{UnruhSonicHole} embodied in Bose-Einstein condensates \cite{BEC,deNovaObservation,Mechanism}, optical systems \cite{optical,fiber,DroriObservation} or classical fluids \cite{classical,WaterTank}, and also inhomogeneous magnets \cite{magnon}, polaritons \cite{polaritons} and Weyl semimetals \cite{Volovik2016,BHinDS,inhomogeneous,StainedSemimetals,fermionic,teemu}. Here we will analyze different Weyl semimetal configurations which can mimic both black and white holes. We motivate idealized model Hamiltonians for these systems. Our models have interesting features, which however do not correspond to conventional expectations for black holes. In particular, the horizon has distinctive local properties, and there is no Hawking radiation.  


{\it Black and White Hole Geometries}

A Weyl semimetal is a material in which the low energy description of the electronic band structure is given by the Hamiltonian $H=\pm \vec{\sigma}  \cdot \vec{p}$, where $\vec{\sigma}$ is the vector of Pauli matrices pertaining to a band index, or spin, $\vec{p}$ is the momentum. For simplicity we set the Fermi velocity to unity. The two signs refer to two chiralities, which generally are located in different places in the Brillouin zone.  Weyl semimetals were  first observed experimentally in tantalum arsenide (TaAs) \cite{weylExp4,weylExp2,weylExp3} and later in a plethora of different materials\cite{WeylReview}. Shortly after, ``type II'' Weyl semimetals \cite{type2}, with over-tilted cones \cite{Bergholtz2015,Xu2015,Trescher2015,Kobayashi2007}, were experimentally observed in e.g. in Molybdenum ditelluride (MoTe$_2$) \cite{expTyp2MoTe}, Tungsten ditelluride (WTe$_2$) \cite{expTyp2WTe2} and in TaIrTe$_4$ \cite{expTyp2TaIrTe}. Additional types were also considered \cite{type3}. Moreover, similar dispersion relations feature in metamaterials such as photonic crystals \cite{PhotonicWeyl} and coupled waveguides \cite{WaveguideWeyl}.

Since condensed matter systems do not have Lorentz symmetry, an extra contribution to the Hamiltonian of the form $H_{\rm tilt} = \vec{\kappa}  \cdot \vec{p}$ can appear, where $ \vec{\kappa}$ is a parameter depending on the details of the material \cite{Bergholtz2015}.  
This term ``tilts" the Weyl cone.  For $\abs{\vec \kappa}>1$ we have a type II dispersion with an over-tipped cone \cite{type2}, and there is only one direction of propagation.  Space-dependent tilting of light cones is of course suggestive of black hole space-times.   To explore this analogy mathematically, we consider the Lagrangian density
\begin{align} \label{lag1}
\mathcal{L} = i \bar \psi \left( \gamma^\mu \partial_\mu + \gamma^0 \vec{\kappa}(\vec{x}) \cdot \vec{\partial } \right) \psi + \text{c.c.}
\end{align}
where $\psi$ is the four component Dirac spinor and $\gamma^\mu$ are the Dirac matrices.  We observe that indeed the Lagrangian density in Eqn. (\ref{lag1}), which describes quasi-particles in a material, is the same as the one for a massless Dirac field in a curved spacetime with the metric 
\begin{equation} \label{metric}
ds^2 = \left( \abs{\vec \kappa}^2 -1 \right) dt^2 - 2 \vec{\kappa} \cdot \vec{dx} dt + \vec{dx} \cdot \vec{dx}.
\end{equation}
In this sense, there is an event horizon at $\abs{\vec \kappa}=1$.

This suggests that a condensed matter system which is described by the Lagrangian in Eqn. (\ref{lag1}) might supply an experimentally accessible vehicle to explore behaviors suggested for black holes, notably including Hawking radiation.   On the other hand, that possibility raises serious issues.  For example, there is no obvious energy source to power continued radiation. Before suggesting how this dilemma can be resolved, let us expand the context of the discussion, by considering the full range of possible ``horizon'' realizations.   To keep things simple, by default we will specialize to one spatial dimension $z$.  As will appear, the incorporation of other additional dimensions can bring in new physical possibilities, which deserve further investigation.

The over-tilted cones imply modes that can propagate only in one direction. The interfaces one can form between a system with supporting such modes and others which support either one-way or two-way propagation are displayed in Table \ref{configurations}.   Since black holes are famous for their power of attraction, at first sight it may seem odd to associate a totally repulsive boundary with a black hole.  But from the point of view of causal structure, the defining characteristic of a black hole horizon is that nothing escapes from the black hole side, interior (whereas its exterior supports both escape and capture).  To achieve that feature, what is essential is that the horizon, viewed from the interior, is totally repulsive. Similarly, the defining feature of a white hole is that it is impossible to throw anything in, and that is achieved through total attraction, viewed from the inside. In Table \ref{configurations}, we show the different possible configurations and their interpretations.   It is also appropriate to note that the materials can terminate at ``end of the world'' boundaries, e.g., boundaries with empty space.  
\begin{table}[b]
\caption{\label{configurations} Interface configuration for type I and type II Weyl semimetal.}

\begin{tabular}{|P{23pt}||P{63pt}|P{63pt}|P{63pt}|}\hline
\diagbox[width=28pt,height=25pt]{L}{R} &\vspace{-30pt} \begin{align} \rightarrow \nonumber \\ \leftarrow \nonumber\end{align} \vspace{-32pt}&\vspace{-30pt} \begin{align} \rightarrow \nonumber \\ \rightarrow \nonumber\end{align} \vspace{-32pt}&\vspace{-30pt} \begin{align} \leftarrow \nonumber \\ \leftarrow \nonumber\end{align} \vspace{-32pt}\\  \hline  \hline
\vspace{-21pt} \begin{align} \rightarrow \nonumber \\ \leftarrow \nonumber\end{align} \vspace{-22pt}  & Normal & Black Hole \newline (On the right) & White Hole \newline (On the right) \\ \hline
\vspace{-21pt} \begin{align} \rightarrow \nonumber \\ \rightarrow \nonumber\end{align} \vspace{-22pt}  & White Hole \newline (On the left)& Chiral \newline matter &Sink \newline junction \\ \hline
\vspace{-21pt} \begin{align} \leftarrow \nonumber \\ \leftarrow \nonumber\end{align} \vspace{-22pt}  &Black Hole \newline (On the left)& Source \newline junction &Chiral \newline matter\\ \hline
  \end{tabular}
\end{table}

{\it Wave Packets and Regularization}

\begin{figure}
\centering
\includegraphics[width=0.48\textwidth]{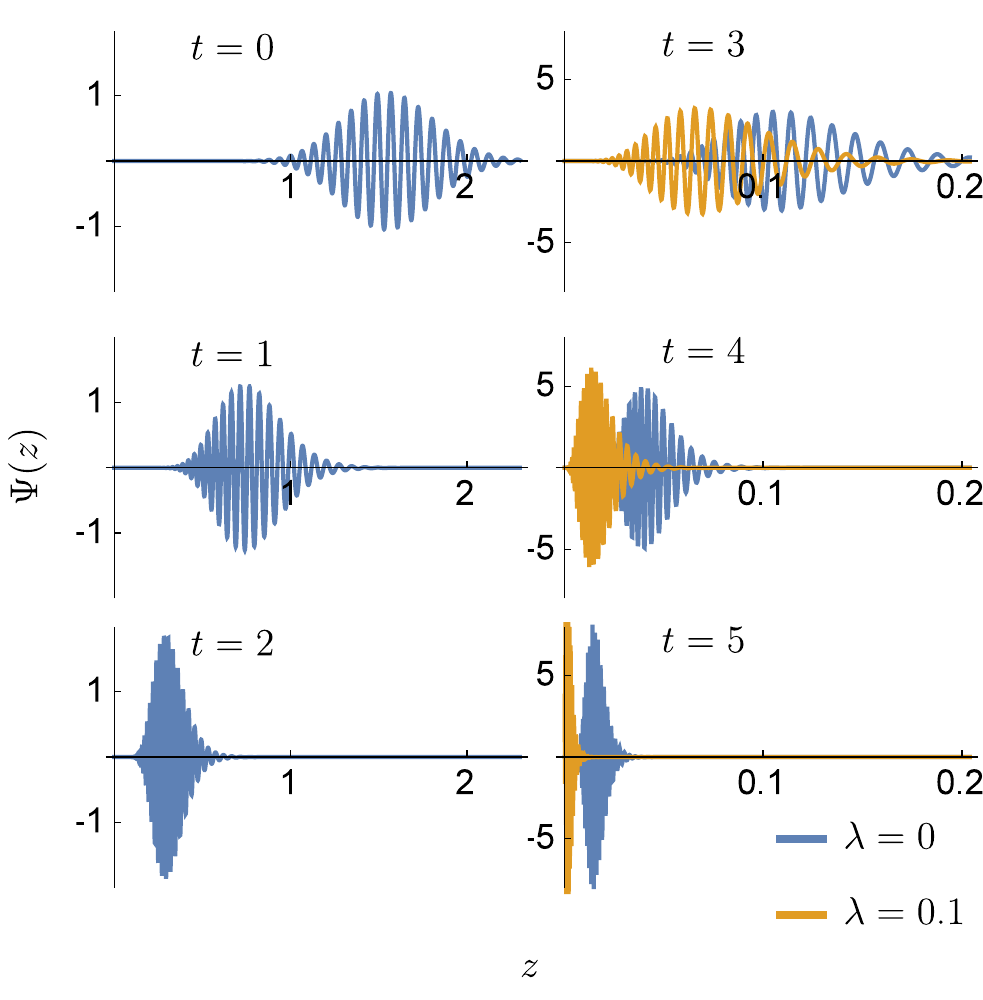}
\caption{The evolution of wavepackets according to Eqn.\,(\ref{characteristic_solution}). Close to the horizon space is compressed, leading to distortion of the waveform. The local amplitude increases, due to the factor $\sqrt{|f (z)|}$ in the denominator, conserving the integrated density. The decreasing velocity ensures that wavepackets supported away from $z=0$ never reach there.  As discussed in the text, with regulators $\lambda>0$ the decrease is less rapid and the solution must be modified.}
\label{wavepackets}
\end{figure}

We can diagonalize $\gamma_z$ and analyze the four components of $\psi$ separately.  This corresponds to isolating definite chirality and spin. We have
 \begin{align} \label{Lag2}
\mathcal{L} = i \psi^* \left[\partial_t +f(z) \partial_z \right] \psi + \text{c.c.}, 
\end{align}   
where $f(z) =  \kappa (z) \pm 1$, with the sign depending on the chirality. The resulting Euler-Lagrange equation
\begin{align} \label{EL1}
\left[\partial_t +f(z) \partial_z + {1 \over 2}{\partial f \over \partial z} \right] \psi = 0
\end{align}
can be solved by the method of characteristics.   Thus, we introduce a variable analogous to the tortoise coordinate of general relativity, 
   \begin{align} \label{tor}
   r(z) =  \int_{z_0}^z { du \over f (u)},
\end{align}
with $z_0$ chosen arbitrarily.  Then we have, formally, the solution
\begin{equation}\label{characteristic_solution}
\psi (t, r) ~=~ \psi_0 (r-t) \sqrt{\frac{|f(r-t)|}{|f(r)|}}
\end{equation}
in terms of a given wave-form $\psi_0 (r)$ at $t=0$.   The subtlety here is that the coordinate $r(z)$ does not necessarily cover all values of $z$.  Indeed, if $f(z)$ is positive for $z > 0$ but approaches $z$ linearly for $z \rightarrow 0$, then the entire line $- \infty < r < \infty$ corresponds to the half-line $0 <  z <  \infty$.  
If we consider wave-packets $\psi_0$ supported on the half-line $0 <  z <  \infty$, then there is no difficulty, and we find the simple qualitative behavior displayed in Fig. \ref{wavepackets}.  Wave packets supported on the half-line $-\infty < z < 0$ evolve analogously.   It appears that the two sides behave independently.  Note that we can also allow finite amplitude or even $1/ \sqrt { |f(z)|}$ singularities at $z=0$, so the evolution of all reasonable initial wave forms is covered.

To quantize the system and construct an appropriate (static) candidate ground state in a conventional way, we must work with energy eigenmodes. Note in this context that there is a preferred time for the electron field, which is set by the external world and ultimately by the fact that after all our solid resides in flat space-time.  More specifically, despite the metric's formal resemblance to black hole solutions of general relativity we need not consider alternative definitions of space-time inspired by the geometry defined by Eqn.\,(\ref{metric}) on an equal footing, nor the ``proper time'' associated with electron trajectories.   The Lagrangian Eqn.\,(\ref{lag1}) from which that geometry was derived is only an approximate model of highly restricted physical situations, and the background material which supports it has no simple generally covariant description.

Solutions of the kind
\begin{align} \label{Sol1}
\psi_\omega (t,z) \propto {e^{-i \omega \left(t -  \int^z { du \over f (u)} \right)} \over \sqrt{\abs{f (z)}}}.
\end{align}
describe the eigenmodes. To use the modes in Eqn. (\ref{Sol1}) in field quantization, we want them to form an orthonormal basis $ \int d\omega \psi_\omega(z_1)  \psi^*_\omega(z_2) = \delta(z_1-z_2), \int dz \psi_\omega(z)  \psi^*_{\omega'}(z) = \delta(\omega-\omega')$.  When $f(z)$ changes sign, say at $z=0$, the integral $\int^{z_1}_{z_2} { du \over f (u)}$ in the exponent can vanish for $z_1\ne z_2$ and modes localized at different positions would not be orthogonal. We resolve this issue by having separate modes in each region, $\psi_\omega^{L} (z)=\psi_\omega (z<0), \psi_\omega^{R} (z)=\psi_\omega (z>0)$.   This separation was to be anticipated, in light of our preceding discussion of the initial value problem.

Another delicacy concerns normalization of the modes.  One typically normalizes by putting the system in a finite sized box.  Here however the effective horizon serves as one side of our box, and it is not entirely clear how to regulate logarithmically divergent normalization integral $\int dz \abs{\psi}^2 \propto dz |f(z)|^{-1}$ (with $f(z) \propto z$).  We can regulate this divergence, and arrive at an unambiguous proposal, by choosing a different form for $f(z)$ near the horizon, {\it viz}.
\begin{align} \label{reg}
  f(z) \propto z {(\lambda + \abs{z})^\alpha \over \abs{z}^\alpha  },
\end{align}
where $\lambda \ge 0$ and $0<\alpha<1$ are parameters.  This represents a steepening of the horizon onset. $\lambda$ sets the length scale for the regularization, while $\alpha$ determines the degree of steepening.  Of course, many other similar regulators could be considered, including regulators which are not (anti)symmetric in $z \rightarrow -z$.

With $f(z)$ of the form Eqn. (\ref{reg}), the tortoise coordinate $r(z)$ converges at the horizon, i.e. the half-line in $z$ maps onto a half-line in $r$.  Following the general (formal) solution Eqn. (\ref{characteristic_solution}), we see that the wave packets now hit the horizon in a finite time.  This, we argue, is a physically reasonable extrapolation of the suggested $\lambda = 0$ behavior.   Indeed, recall that in that limit we had wave packets accumulating at the horizon, yet never quite reaching it.  But our solid has a natural minimum length scale, i.e. the lattice spacing, and once the packet is predicted to be squeezed below that scale, it should have arrived.   That heuristic argument helps make the regulator appear physically reasonable, but it also foretells the breakdown of the formal solution Eqn. (\ref{characteristic_solution}).  To proceed further, we must extend the solution in a way consistent with physical requirements.  

\begin{figure}
\centering\includegraphics[trim=1cm 2.8cm 1cm 18.5cm, clip=true,width=0.48\textwidth]{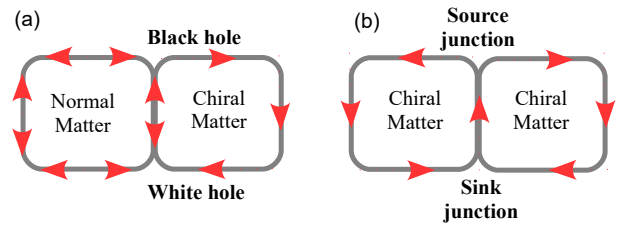}
\caption{Interfaces of chiral matter. The Lagrangian in Eqn. (\ref{Lag2}) describes edge states of a chiral bulk, where the sign of $f(z)$ determines the chirality. The change of sign at the horizon is where edge states of two materials with different chiralities meet, with the interpretation given by Table \ref{configurations}. The propagation along the interface is the analogue of a wormhole connecting a black hole to a white one.}
\label{edgeModes}
\end{figure}

There are two possibilities which suggest themselves immediately, but do not withstand scrutiny.  We might try to allow the wave-packet to reflect back to where it came from.  But this is impossible, because the propagation is unidirectional.  Or, we might try to allow the wave-packet to emerge on the opposite side of the horizon.  But solutions on the other side propagate back toward the horizon, so no escape is possible.  Thus it appears we must allow the wave packet to accumulate on the horizon, or to propagate elsewhere.  The latter possibility can arise for chiral edge modes that bound higher-dimensional regions, as indicated in Fig. \ref{edgeModes}.   The latter case can certainly arise in the context of the quantum Hall effect \cite{Stone} and plausibly in other situations.  Here, however, we shall pursue the former.   It is similar in spirit to the ``membrane paradigm'' of black hole physics \cite{membrane}, though in our context the membrane represents true local physics, and its properties are not so tightly constrained, as there is no analogue of the ``no hair'' theorem.

{\it Model Hamiltonian and Qualitative Behavior}

The preceding considerations suggest, in the context of quantization, that we should add a degree of freedom localized on the horizon which couples to the excitations described by the regularized eigenmodes.  Let us consider how this can work.


The Hamiltonian for the regularized eigenmodes is given by 
 \begin{align} \label{H0}
 H_0= \sum_{i=R,L} \int^\infty_0 d\omega  \omega \left( a^{\dagger}_{i,\omega} a_{i,\omega} +  b^{\dagger}_{i,\omega} b_{i,\omega}\right),
\end{align}
where $a_{i,\omega}^{(\dagger)}$ a (creation) annihilation operator for the mode $\psi^{i}_\omega$ and we relabeled $a^{\dagger}_{i,\omega<0} \rightarrow b_{i,-\omega}, a_{i,\omega<0} \rightarrow b^{\dagger}_{i,-\omega}$. The coupling to the localized degree of freedom is given by
\begin{align} \label{HI}
H_I = \sum_{i=R,L} \int^\infty_{-\infty} d\omega g_i( \omega)   \left(A a^{\dagger}_{i,\omega}  +  A^{\dagger} a_{i,\omega}\right),
\end{align} 
where $A^{(\dagger)}$ is a (raising) lowering operator for the charge on the horizon, $Q=- A^{\dagger}A$, and $g_i( \omega)$ is a coupling function such that $g_i( z) = \int d\omega \psi^{i}_\omega(z) g_i( \omega)$ have significant support only close to the horizon. The energetic cost of charge accumulation is given by an onsite Hamiltonian $H_c = E_c Q^2$, with the charging energy $E_c$ depending on microscopic details. Diagonalizing the complete Hamiltonian $H=H_0 + H_I + H_c$, the regularized eigenmodes are slightly modified, or dressed. Since the integral in Eqn. (\ref{HI}) includes negative frequencies, we have terms such as $A^{\dagger} b^{\dagger}_{i,\omega}$ and $A b_{i,\omega}$. Thus if a particle reaching the horizon decreases the charge, a hole increases it.

\begin{figure}
\centering\includegraphics[trim=5.5cm 0.3cm 10cm 2.7cm, clip=true,width=0.48\textwidth]{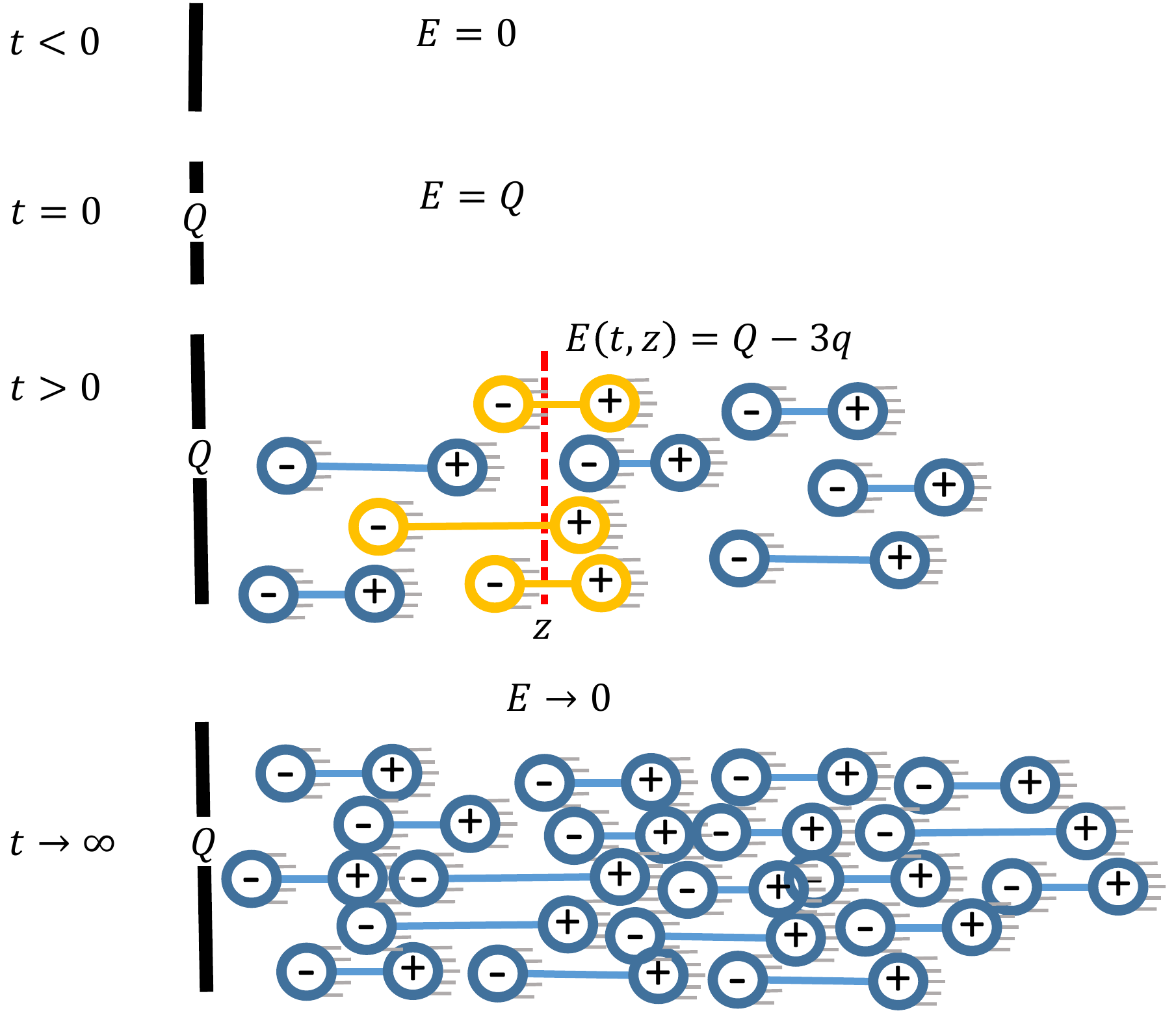}
\caption{Screening of charge in a type II Weyl semimetal. The pairs created by the Schwinger mechanism will propagate to one side. As long as the electric field persist more pairs will be created. The study state is a polarized bulk with vanishing electric field.}
\label{pairCreation}
\end{figure}

Let us now look on the response of the system to a few types of external perturbations. First, consider applying a voltage across the system, i.e. running a current. We start with the case of a white hole. The external potential will create positively charged excitations, i.e. holes, on one side and negatively charged
ones on the other. Both excitations will propagate towards the interface, canceling each others charge. To the experimentalist measuring the current it would simply look as if charge has moved from one side to the other. (Microscopic details might lead to a finite resistance or to local intermediate charge accumulation, which is the case in any inhomogeneous system). The case of a black hole is analogous, but the excitations are created at the interface and propagate to the two sides.  This mechanism is reminiscent of the picture of Hawking radiation, based on pair creation \cite{pairEmission,tunneling}, but the details are quite different.

We can consider another scenario of placing some charge at the interface, say in a quenched way. In a type I Weyl semimetal, particle-hole pairs created by the Schwinger mechanism will screen the charge. A similar process occurs for type II as well. In contrast to the usual case, where the particle and the hole move in opposite direction due to the electric field, here both move in the same direction. However, as shown in Fig. \ref{pairCreation}, this phenomenon still allows for screening to occur. 

\begin{figure}
\centering\includegraphics[trim=2cm 3cm 12cm 1cm, clip=true,width=0.5\textwidth]{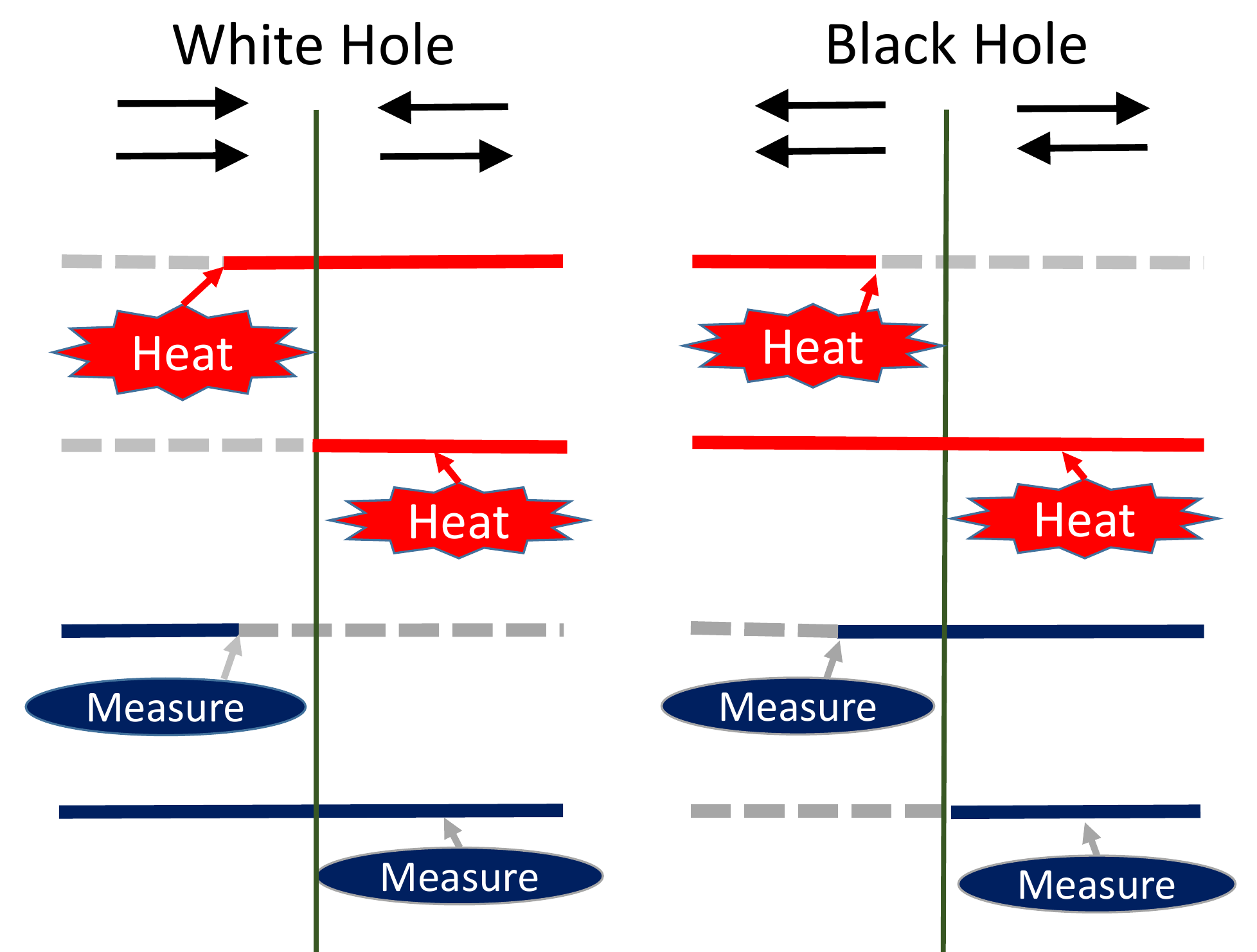}
\caption{Casual structure of an type I-type II Weyl semimetals interface. The colored region are accessible to the relevant operation, heating or measuring.  Time reversal interchanges those operations.}
\label{heatFlow}
\end{figure}

An interesting feature of the causal structure emerges if we analyze the heat transfer in this system. Unlike charge for which an excitation can be negative or positive, all excitations imply adding energy to the system. In a system that has only left moving modes, heating a particular location will not affect anything to the right of that location. For any interface configuration, it is straightforward to see which region can can be heated from any position and from which location one can measure any region. This is shown in Fig. \ref{heatFlow}.   We see that the time-reversal exchange of black and white holes involves interchange of the excitation and measurement processes.  

{\it Summary and Discussion}

We have demonstrated that models of fermion propagation which arise as simple idealizations of behavior in type II Weyl semimetals and their junctions map onto models of fermion propagation in space-times with unusual causal structures: black holes, white holes, universal sources and universal sinks.  We analyzed the behavior of wave packets and eigenmodes in those contexts, suggested how the models can be quantized in a way that regulates the infinite propagation delays at the horizon, at the cost of introducing an explicit boundary degree of freedom.  We analyzed some simple but characteristic situations qualitatively, to demonstrate the physical consistency of this modeling.  

In realistic Weyl semimetals there are always additional states at the Fermi level, in addition to those described by the linearized low energy theory with a tilted Weyl cone. Most materials host parasitic bands that cross the Fermi level, and usually the Weyl cones occur at finite doping. Those are however not fundamental problems, and there are materials where they are avoided \cite{IdealWeyl}. There are however also complications of a more fundamental nature: (i) The Nielsen Ninomiya theorem dictates that there are an even number of Weyl points in the Brillouin zone; (ii) The Weyl points are connected by Fermi arcs on the surface; (iii) The ``light cones'' are only approximate - to recover periodicity in the BZ the bands must curve back down (or up) away from the Weyl point.  Those complications do not necessarily render the physical picture suggested by our linearized model of a single Weyl node with varying tilt irrelevant, because: (i) different Weyl nodes are essentially independent, at least in materials where they are well separated in momentum space; (ii) for appropriate surfaces, the fermi arcs can be made small;  (iii) band curvature introduces quantitative, but not qualitative alteration of the model.  Prospects for engineering materials with the relevant properties as discussed in \cite{BurkovBalents}.

Although the starting equations were the same, the physical picture we have arrived at for our material system is quite different from conventional expectations for black holes.   Among other things, our regularized horizon corresponds to a naked singularity, there is no Hawking radiation, and other types of exotic backgrounds (white holes, sources, sinks) appear on an equal footing with black holes.  How did this happen, and what does it mean? The key point is that the equations of motion for some chosen degrees of freedom do not fully determine, independent of context, which solutions are physically relevant, nor how they should be interpreted. In our case, the existence of a preferred time and of a locally identifiable horizon are salient features of our material system which directly contradict standard, well-motivated assumptions about black holes.   More generally, all proposed material analogues of black hole behavior require critical scrutiny, to determine how far the analogy can be taken.  That said, consideration of material systems with unusual properties is valuable in itself, and might suggest new possibilities for the material system we call space-time.

\textit{Acknowledgement}: We gratefully acknowledge useful discussions with S\"oren Holst, Teemu Ojanen and Kostya Zarembo. This work was supported by the Swedish Research Council under Contracts No. 335-2014-7424, 2016-04192$\_$3, 2018-06720$\_$3  and the Wallenberg Academy Fellows program of the Knut and Alice Wallenberg Foundation. In addition, FW's work is supported by the U.S. Department of Energy under grant Contract  No. DE-SC0012567 and by the European Research Council under grant 742104.

\bibliography{NoHawkingLib}

\end{document}